%% file: interspeech_v11.tex
\documentclass[a4paper]{article}
\usepackage{INTERSPEECH2022}
\usepackage{amsmath}

\usepackage[ruled,linesnumbered]{algorithm2e}
\usepackage{hyperref}
\usepackage{color}
\usepackage{subfigure}
\title{Rainbow Keywords: Efficient Incremental Learning for Online Spoken Keyword Spotting}
\name{Yang Xiao, \thanks{* Nana Hou contributed to this work before leaving Nanyang Technological University, Singapore.}Nana Hou$^*$, Eng Siong Chng}
\address{
  School of Computer Science and Engineering, Nanyang Technological University, Singapore.}
\email{\{yxiao009,nana001\}@e.ntu.edu.sg,aseschng@ntu.edu.sg}

\makeatletter
\def\blfootnote{\xdef\@thefnmark{}\@footnotetext}
\makeatother

\begin{document}
% \ninept
\maketitle
\begin{abstract}
Catastrophic forgetting is a thorny challenge when updating keyword spotting (KWS) models after deployment. This problem will be more challenging if KWS models are further required for edge devices due to their limited memory. To alleviate such an issue, we propose a novel diversity-aware incremental learning method named Rainbow Keywords (RK). Specifically, the proposed RK approach introduces a diversity-aware sampler to select a diverse set from historical and incoming keywords by calculating classification uncertainty. As a result, the RK approach can incrementally learn new tasks without forgetting prior knowledge. Besides, the RK approach also proposes data augmentation and knowledge distillation loss function for efficient memory management on the edge device. Experimental results show that the proposed RK approach achieves 4.2\% absolute improvement in terms of average accuracy over the best baseline on \textit{Google Speech Command} dataset with less required memory. The scripts are available on GitHub \footnote{https://github.com/swagshaw/Rainbow-Keywords}.
\end{abstract}

\section{Introduction}
% Background: Spoken keyword spotting (KWS) are widely utilized on the edge device.
Spoken keyword spotting (KWS) \cite{lopezespejo2021deep} aims to identify the specific keywords in the audio input. It serves as a primary module in many real-world applications, such as Apple Siri and Google Home, which are widely utilized on the edge device \cite{zhang2018hello,hz3}. Current deep-learning-based keyword spotting systems are usually trained with limited keywords in the compact model for lower computation and smaller footprint \cite{chen2014kws,berg2021kwt,ng2022convmixer,kim2021broadcasted}. Therefore, the performance of the KWS model trained by the source-domain data may degrade significantly when confronted with unseen keywords of the target-domain at run-time \cite{nana2,nana5}. 
% \vspace*{1\baselineskip} 

To alleviate such a problem, prior work \cite{awasthi2021teaching,2021Few} utilize few-shot fine-tuning \cite{hz5} to adapt KWS models with training data from the target-domain for new scenarios. However, performances on data from the source domain after adaptation could be poor, which is also known as the  \textit{catastrophic forgetting} problem \cite{mccloskey1989catastrophic}. Recent work \cite{pclkws2022} proposes a progressive continual learning \cite{delange2021continual,hz2,voraefficient} strategy for small-footprint keyword spotting to alleviate the catastrophic forgetting problem when adapting the model trained by source-domain data with the target-domain data. The limitations of such an approach are two-fold. First, the approach requires the task-ID as auxiliary information to learn the knowledge of different tasks, which is not always available in practice. Second, the storage volume occupied by the model will increase with the higher task numbers \cite{nana3}. The storage volume will be unaffordable for light edge devices. 
% \vspace*{1\baselineskip} 

This paper proposes a novel diversity-aware incremental learning approach named Rainbow Keywords (RK) to address the issues mentioned above, requiring no task-ID information with fewer parameters. Specifically, the proposed RK approach introduces a diversity-aware sampler to select few but diverse examples from historical and incoming keywords by calculating classification uncertainty. As a result, the model will not forget the prior knowledge when learning new keywords even utilizing limited historical examples. Furthermore, we utilize a mixed-labeled data augmentation to additionally improve the diversity of selected examples for higher performances. Besides, we propose a knowledge distillation loss function to guarantee that the prior knowledge could remain from the limited selected examples. We conduct our experiments on \textit{Google Speech Command} dataset following the setup of prior work \cite{mai2022online,prabhu2020gdumb}. Experimental results show that the proposed RK approach achieves 4.2\% absolute improvement in terms of Average Accuracy over the best baseline with less required memory.

\section{RK Architecture}
With online keyword spotting systems, we assume that the model should identify all keywords in a series of tasks without catastrophic forgetting. For each task $\tau_t$, we have input pairs $(x_t,y_t)$, where $x_t$ denote audio utterances and $y_t$ are keyword labels. We aim to minimize a cross-entropy loss \cite{masana2020class,hz1} of all keywords $N^t$ up to the current task $\tau_t$ formulated as :
\begin{equation}
L_{CE}(x,y) = \sum\limits^{N^t}_{i=1} y_i log \frac{exp(o_i)}{\sum^{N^t}_{j=1} exp(o_j)}
\label{eq1}
\end{equation}
where $o$ denotes the output logits of the model in the task $\tau_t$.

\subsection{Rainbow Keywords Network}
% Link of Drawio: https://drive.google.com/file/d/1LXiLt0TItwgPmuYkIeu-ucaM63LRNEzi/view?usp=sharing
\begin{figure*}[t]
  \centering
  \includegraphics[width=0.9\linewidth, height=0.556\linewidth]{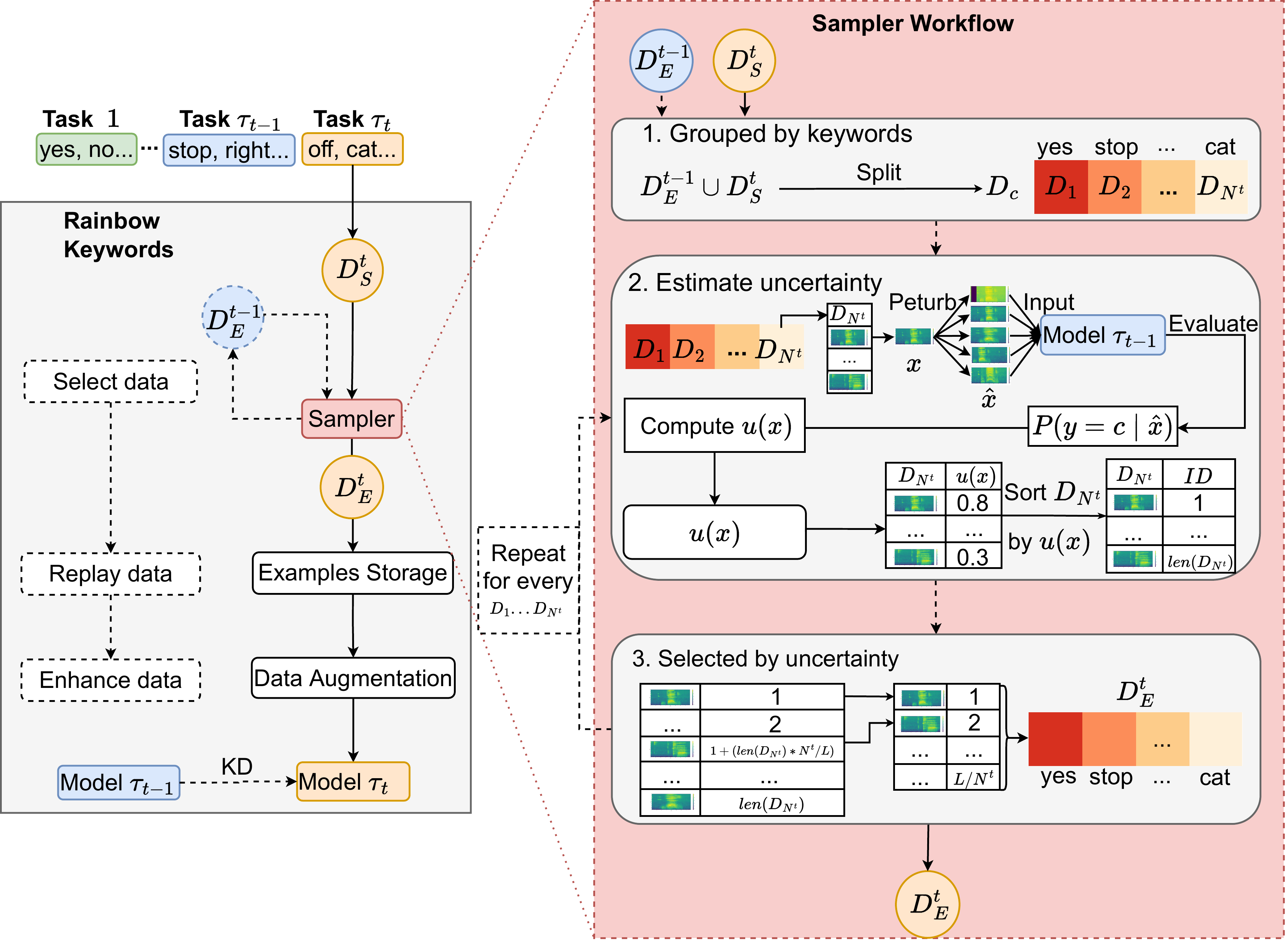}
  \caption{Block diagram of the proposed Rainbow Keywords approach. Specifically, $D^t_S$ denotes incoming audio stream data of the task $\tau_t$. $D^t_E$ and $D^{t-1}_E$ denote the examples of the task $\tau_t$ and $\tau_{t-1}$, respectively. We group $D^{t-1}_E \cup D^t_S$ into subsets as $D_c, c=1...N^t$ by unique keywords, where $N^t$ denotes the total numbers of unique keywords in $D^{t-1}_E \cup D^t_S$ set. $x$, $\hat{x}$ and $K$ present each sample in $D_c$, the five perturbations of $x$ and the five perturbation strategies. ``Compute $u(x)$" is to compute $u(x)$ by Eq.\ref{ux}.}
  \label{fig:workflow}
\end{figure*}

We now introduce the proposed Rainbow Keywords (RK) approach, which consists of three parts: a diversity-aware sampler, a data augmentation strategy, and a knowledge distillation loss function. The detailed workflow of each module is as followed.

As illustrated in Figure \ref{fig:workflow}, we first receive several incoming audio utterances of the task $\tau_t$ as inputs $D^t_S$, which are fed to the sampler together with historical examples $D^{t-1}_E$. Next, the diversity-aware sampler selects diverse examples $D^t_E$ for task $\tau_t$ by calculating classification uncertainty. Such selected examples $D^t_E$ are then saved in examples storage and further augmented by the data augmentation to increase their diversity. Finally, we copy the old model as the teacher model and update the student model with augmented examples $D^t_E$ with the knowledge distillation loss function.

\subsubsection{Diversity-Aware Sampler}
The diversity-aware sampler aims to select \textit{diverse examples} to manage memory efficiently. Such diverse examples are defined by the relative location of each example in feature space, which are estimated by the uncertainty of the sample through the inference by the classification model \cite{bang2021rainbow}. The required three steps are shown in the right red box of Figure \ref{fig:workflow}: 

% \vspace*{1\baselineskip} 

\textbf{Split by keywords:} The first step gathers the historical examples $D^{t-1}_E$ and incoming data $D^t_S$, and groups them into subsets as $D_c, c=1...N^t$ by unique keywords, where $N^t$ denotes the total numbers of unique keywords in $D^{t-1}_E \cup D^t_S$ set. 

% \vspace*{1\baselineskip} 

\textbf{Estimate uncertainty:} The second step estimates the uncertainty of each sample $x$ in $D_c$ by Monte-Carlo (MC) method \cite{gal2016dropout}, which is defined in Equation \ref{eq3}. 
\begin{equation}
\label{eq3}
    P(y = c \mid x) = \int p(y=c \mid \hat{x})p(\hat{x}\mid x) d\hat{x}
\end{equation}
where $x$, $\hat{x}$, $y$ denote each audio utterance of keyword $D_{N^t}$, the five perturbations of $x$, and the label of $x$. Therefore, the uncertainty of the audio utterance $x$ is formulated as $u(x)$: 
\begin{equation}
\label{ux}
    u(x) \approx 1- \frac{1}{K} \sum\limits^K_{k=1} P(y = c\mid \hat{x}_k)
\end{equation}

where $K$ presents five perturbation strategies, including Clipping Distortion \cite{park2019specaugment}, TimeMask \cite{park2019specaugment}, Shift \cite{ko2015audio}, PitchShift \cite{ko2015audio} and FrequencyMask \cite{park2019specaugment}. The larger $u(x)$ indicates that the less confidence of model to predict the perturbations.
% \vspace*{1\baselineskip} 

\textbf{Select by uncertainty:} The third step selects $L$ examples from $D_c$ descending by uncertainty $u(x)$ with the step size of $len(D_c)*N^t/L$. As a result, the most diversity examples are included in $D_{E}^{t}$. Only these examples are available for training.

\subsubsection{Data Augmentation}
As the examples in $D_{E}^{t}$ are few due to memory limitation, we apply the data augmentation to further increase the diversity \cite{nana1} of $D_{E}^{t}$. Specifically, we randomly mix two audio utterances to increase the amounts of training data without extra storage.

\subsubsection{Knowledge Distillation Loss}
Recent studies \cite{hinton2015distilling, wu2019bic,rebuffi2017icarl} show that Knowledge Distillation (KD) is effective for transferring knowledge between teacher-student models. Inspired by such theory, we consider the model of task $\tau_{t-1}$ as the teacher model and the model of task $\tau_{t}$ as the student model. We propose a knowledge distillation loss to preserve the prior knowledge from the teacher for the student model to avoid catastrophic forgetting, which is formulated as:

 % 详细讲一讲know distillation
\begin{equation}
\begin{aligned}
\sigma(o_i^t(x);N^{t-1}) & = \frac{exp(o_i^t/T)}{\sum^{N^{t-1}}_{j=1} exp(o_j^t/T)} \\
L_{KD}(o^t(x),o^{t-1}(x)) & = \sum\limits^{N^{t-1}}_{i=1} \sigma(o^{t}_i(x)) log \sigma(o^{t-1}_i(x))
\label{eq7}
\end{aligned}
\end{equation}
where $o^{t-1}(x)$ and $o^t(x)$ denote the output logits of the teacher model and student model, respectively. $N^t$ is all keywords up to the task $\tau_t$. $\sigma$ is the knowledge distillation softmax function parameterized by the temperature $T$. The temperature $T$ is the experiential hyper-parameters of knowledge distillation set as 2.0. As a result, we aim to minimize the total loss of all keywords $N^t$ up to the current task $\tau_{t}$ formulated as:
\begin{equation}
\begin{aligned}
L_{total}(x,y) & = \lambda L_{CE}(x,y) + \\ & (1-\lambda) L_{KD}(o^t(x),o^{t-1}(x))
\end{aligned}
\end{equation}
Where $L_{CE}$ is the cross-entropy loss defined in Eq.\ref{eq1}, and $L_{KD}$ is the knowledge distillation loss defined above. $\lambda$ is the experiential hyper-parameters defined as $\sqrt{1-\frac{N^{t-1}}{N^t}}$.

\begin{figure*}[t]
  \centering
  \begin{minipage}[t]{0.24\linewidth}
  \centering
  \includegraphics[width=\linewidth]{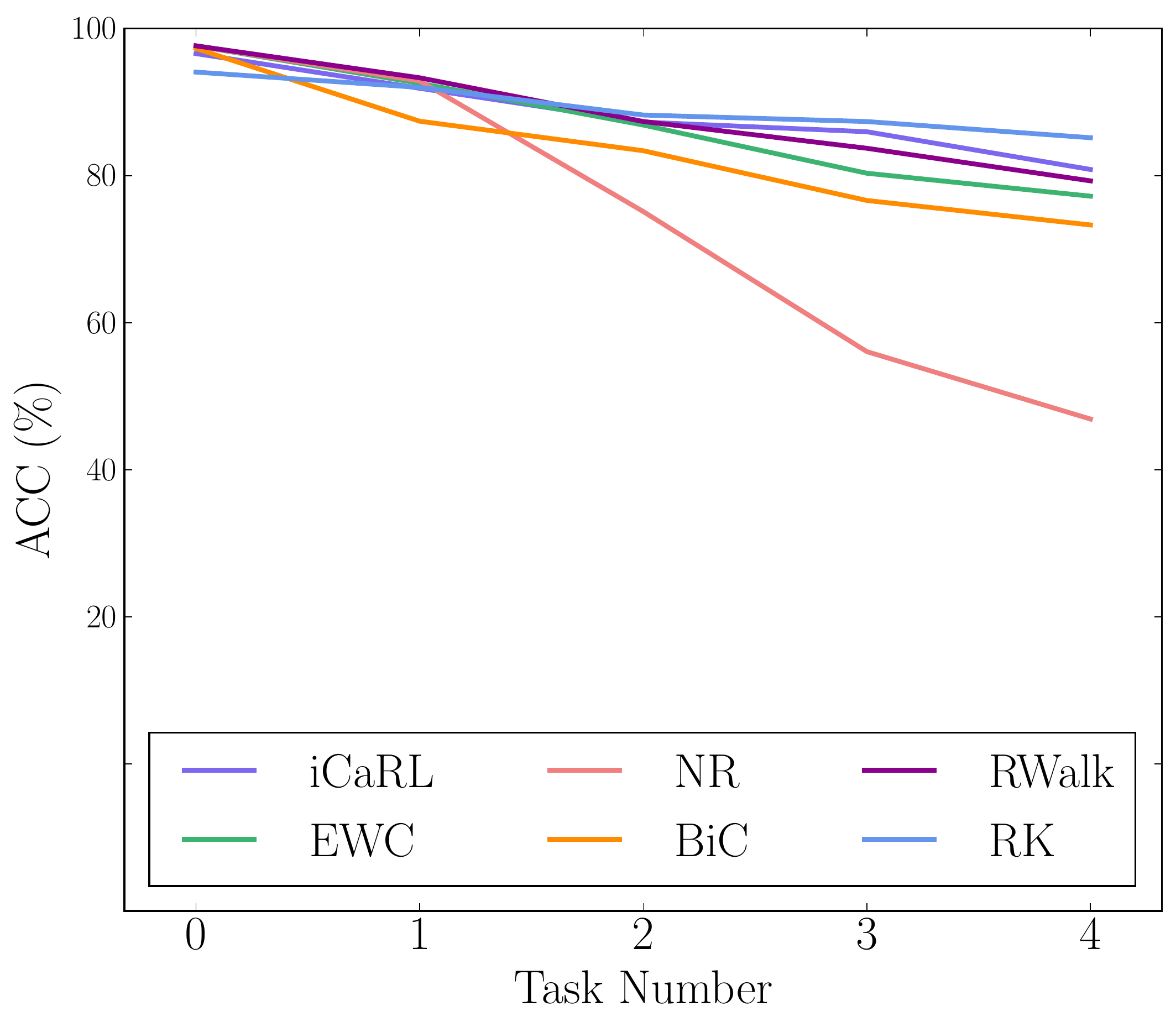}
  \centerline{(a)}
  \end{minipage}
\hfill
  \begin{minipage}[t]{0.24\linewidth}
  \centering
    \includegraphics[width=\linewidth]{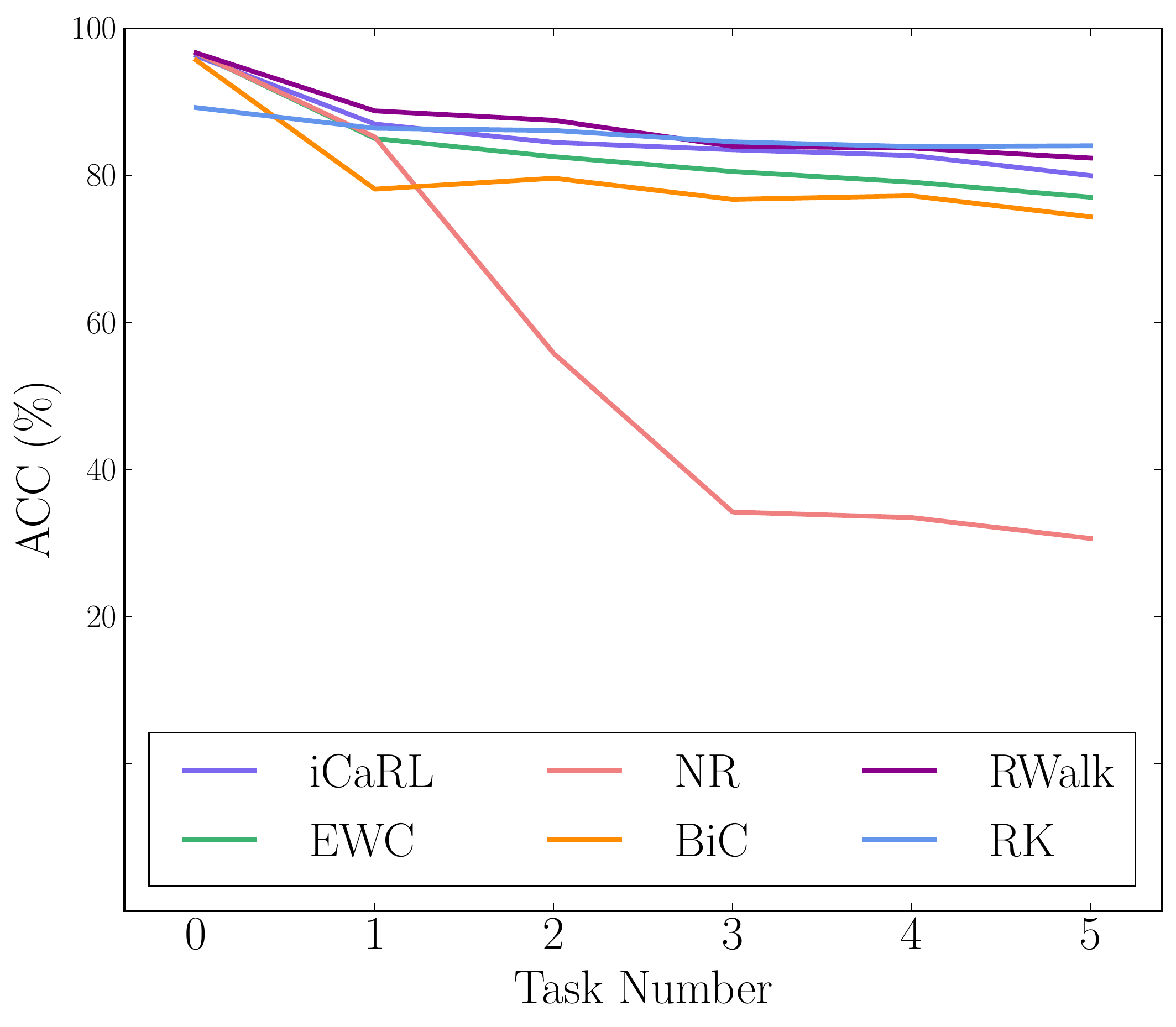}
  \centerline{(b)}    
  \end{minipage}
\hfill
  \begin{minipage}[t]{0.24\linewidth}
  \centering
  \includegraphics[width=\linewidth]{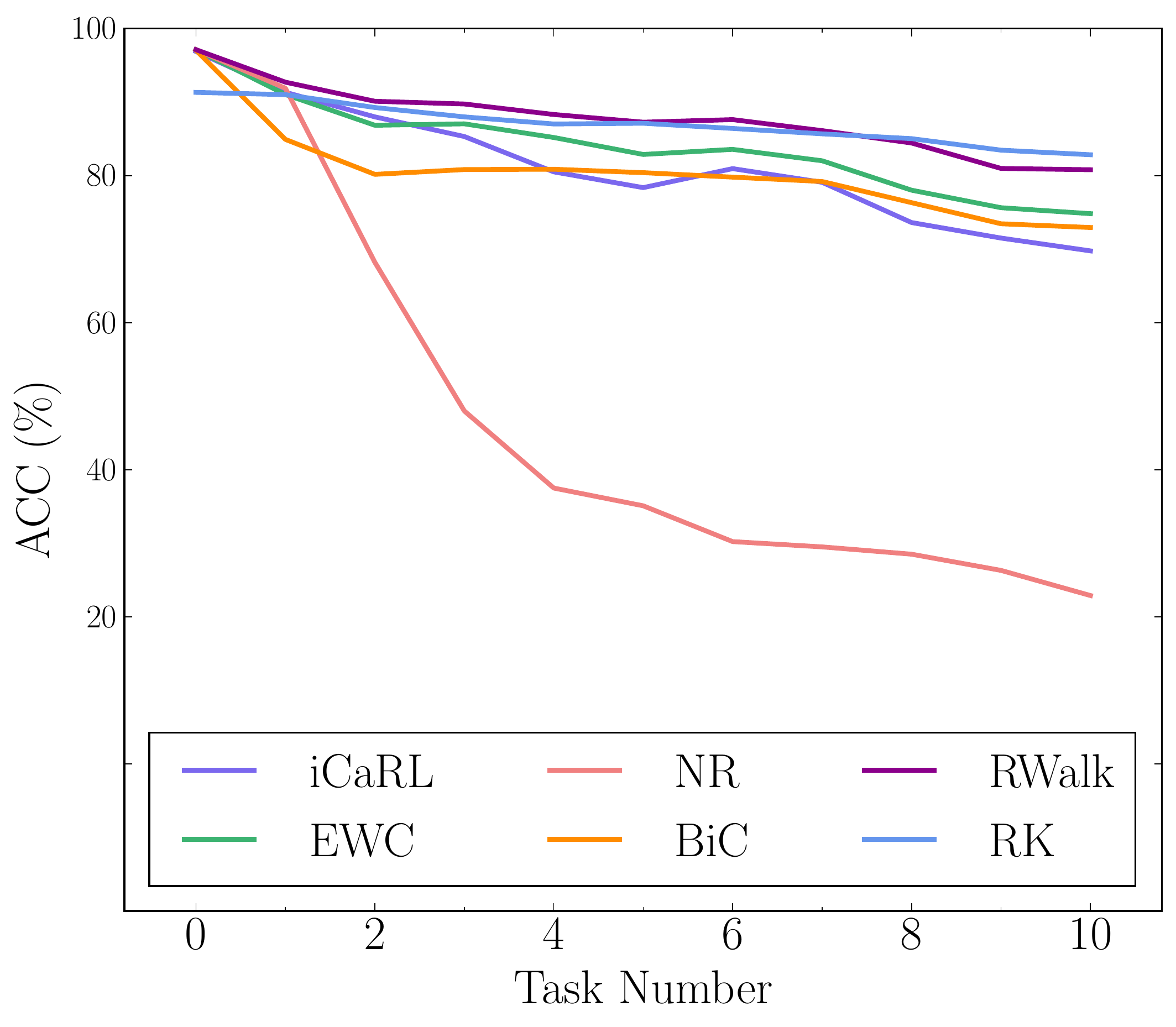}
  \centerline{(c)}
  \end{minipage}
\hfill
  \begin{minipage}[t]{0.24\linewidth}
  \centering
  \includegraphics[width=\linewidth]{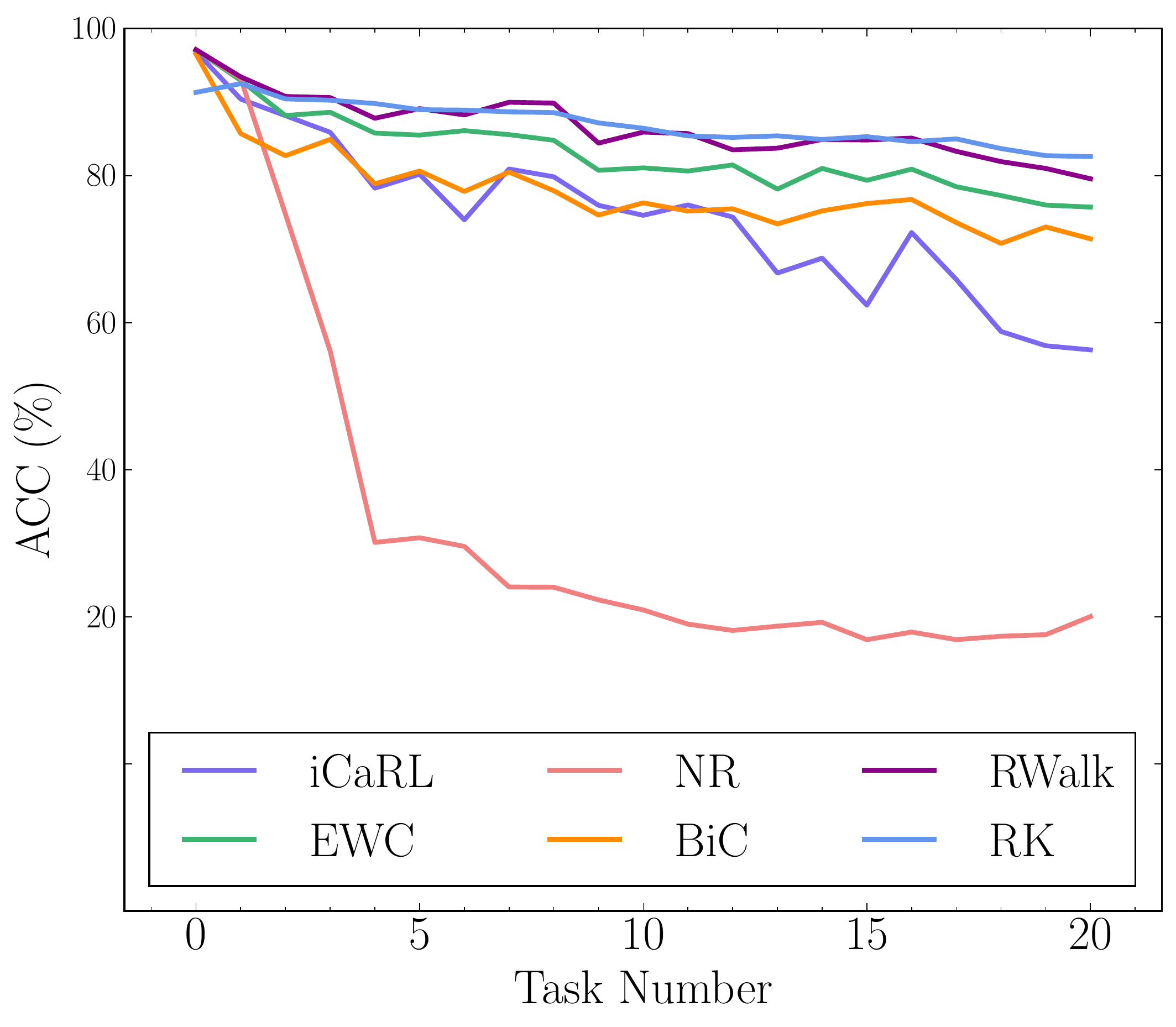}
  \centerline{(d)}
  \end{minipage}
  \caption{The ACC (\%) in a comparative study of increasing task numbers on the proposed RK approach and other competitive baselines. Figures from (a) to (d) represent the experiment with task numbers (= 20, 10, 5, 4).}
  \label{fig:task}
\end{figure*}

\section{Experiments and Results}
\subsection{Dataset}
We conduct experiments on the \textit{Google Speech Command}
dataset v1 (GSC) \cite{warden2018speech}, which includes 64,727 one-second audio clips with 30 English keywords categories. We utilize 80\% of data for training and 20\% of data for testing. All of the audio clips in GSC are sampled at 16kHz in our experiment. 

\subsection{Experimental Setup}
\subsubsection{Network configuration} 
We employ the TC-ResNet-8 \cite{choi2019temporal} as our testbed to evaluate the proposed rainbow keywords approach. It includes a 1-D convolution layer followed by three residual blocks, which consist of 1-D convolution, batch normalization and ReLU active function. Each layer has \{16,24,32,48\} channels.

% \vspace*{1\baselineskip} 

We first pre-train the TC-ResNet-8 model on the GSC dataset with 32,000 audio clips, including 15 unique keywords. To evaluate the learning ability of the proposed RK approach, we split the rest data as 5 tasks. Each task includes 3 new unique keywords, which is unseen in previous tasks. To simulate the condition of edge devices, we set the max amount of examples $L$ due to the limited memory in edge devices \cite{mai2022online,prabhu2020gdumb}. 

% \vspace*{1\baselineskip} 

During the training stage, we utilize the Mel-frequency cepstrum coefficients (MFCC = 40) as inputs. The network is optimized by the Adam \cite{kingma2014adam} algorithm with the learning rate 0.1. The batch size is set to 128 and the number of epochs is 50. 

\subsubsection{Reference baselines}
% 每一个reference三句话简单讲一讲
We built eight baselines for comparisons. 
% \vspace*{1\baselineskip} 
\begin{itemize}
    \item \textbf{Fine-tune training}: adapts the TC-ResNet-8 model for each new task without class incremental learning (CIL) strategies. We consider it as the lower-bound baseline.
    \item \textbf{NR \cite{hsu2018nr}:} is a CIL approach which randomly selects training samples from previous tasks for future training.
    \item \textbf{iCaRL \cite{rebuffi2017icarl}:} is a CIL approach which selects the samples close to the mean of its own class. Then iCaRL utilizes examples for future training.
    \item \textbf{EWC \cite{ewc}:} is a CIL approach which incorporates a quadratic penalty to regularize parameters of model that were important to past tasks. The importance of parameters is approximated by the Fisher Information Matrix.
    \item \textbf{RWalk \cite{rwalk}:} is a CIL approach which improves the EWC. Both Fisher Information Matrix approximation \cite{ewc} and online path integral \cite{zenke2017pi} are fused to calculate the importance for each parameter. RWalk also selects historical examples and utilizes them for future training.
    \item \textbf{BiC \cite{wu2019bic}:} is a recent CIL approach with more attentions. BiC introduces an additional layer to correct task bias of the network. BiC also uses the same sampling method as iCaRL to select historical examples.
    \item \textbf{PCL-KWS \cite{pclkws2022}:} is a CL approach for Spoken Keyword Spotting. Specifically, the PCL-KWS includes several task-specific sub-networks to memorize the knowledge of the previous keywords. Then, a keyword-aware network scaling mechanism is introduced to reduce the network parameters. The PCL-KWS requires the task-ID to select conspronding sub-networks.
    \item \textbf{Joint training:} trains the TC-ResNet-8 model with the whole dataset, regardless of any constrains. We consider it as the upper-bound baseline.
\end{itemize}

\subsubsection{Metrics}
% Metric 选择KWS常用的mertic
We report performances in terms of the accuracy and efficiency metrics. The accuracy metrics include \textit{Average Accuracy} (ACC), and \textit{Backward Transfer} (BWT) \cite{lopez2017gradient}. Specifically, the `Average Accuracy' reports an accuracy averaged on all learned tasks after the entire training ends. The “BWT” evaluates accuracy changes on all previous tasks after learning a new task, indicating the forgetting degree. The efficiency metrics include Parameters and Memory \cite{nana4}. The `Parameter' measures the total parameters of the model in the strategy. The `Memory' indicates the memory requirement of total training data in each task.

\subsection{Results}
% 后面的单词换小写
\begin{table}[tbp]
\centering
\caption{Average Accuracy (ACC) and Backward Transfer (BWT) in a comparative study of the proposed KD Loss. $L$ denotes the memory size for RK.}
\label{tab:kd}
\resizebox{0.4\textwidth}{!}{%
\begin{tabular}{l|c|c|c}
\hline
\hline
\textbf{Methods(L=500)} & \textbf{KD Loss} & \textbf{ACC($\uparrow$)} & \textbf{BWT($\uparrow$)} \\ \hline\hline
Rainbow Keywords & NO  & 0.779 & -0.033 \\ \hline
Rainbow Keywords & YES  & \textbf{0.779}         & \textbf{-0.015}                   \\ \hline\hline
\end{tabular}%
}
\end{table}
\subsubsection{Effect of the knowledge distillation loss}
We first analyse and summarize the performances with knowledge distillation loss. As shown in Table \ref{tab:kd}, we observe that the proposed knowledge distillation loss function achieves 54.5\% relative improvements in terms of BWT.

\subsubsection{Effect of various task numbers on RK approach}
We further analyse and summarize the performances of the proposed RK approach with increasing task numbers as shown in Figure \ref{fig:task}. The x-axis represents the task numbers (= 20, 10, 5, 4) and the y-axis is the evaluation metric of ACC. For example, when the task number is set to 20, we pre-train the TC-ResNet-8 with 21,000 audio clips including 10 unique keywords and the rest data is split into 20 tasks including 1 unique unseen keyword. The corresponding ACC is the accuracy of the testing set after each task is finished training. The memory size $L$ is 1500 for the proposed RK approach here. We observe that the proposed RK approach achieves the best ACC performances with increasing task numbers. Even though the difficulty of preserving prior knowledge is increased with the increasing task numbers, the proposed RK approach still can obtain over 85.0\% ACC, which is much better than other baselines.

\begin{figure}[!t]
   \centering
  \begin{minipage}[tbp]{0.49\linewidth}
  \centering
  \includegraphics[width=\linewidth]{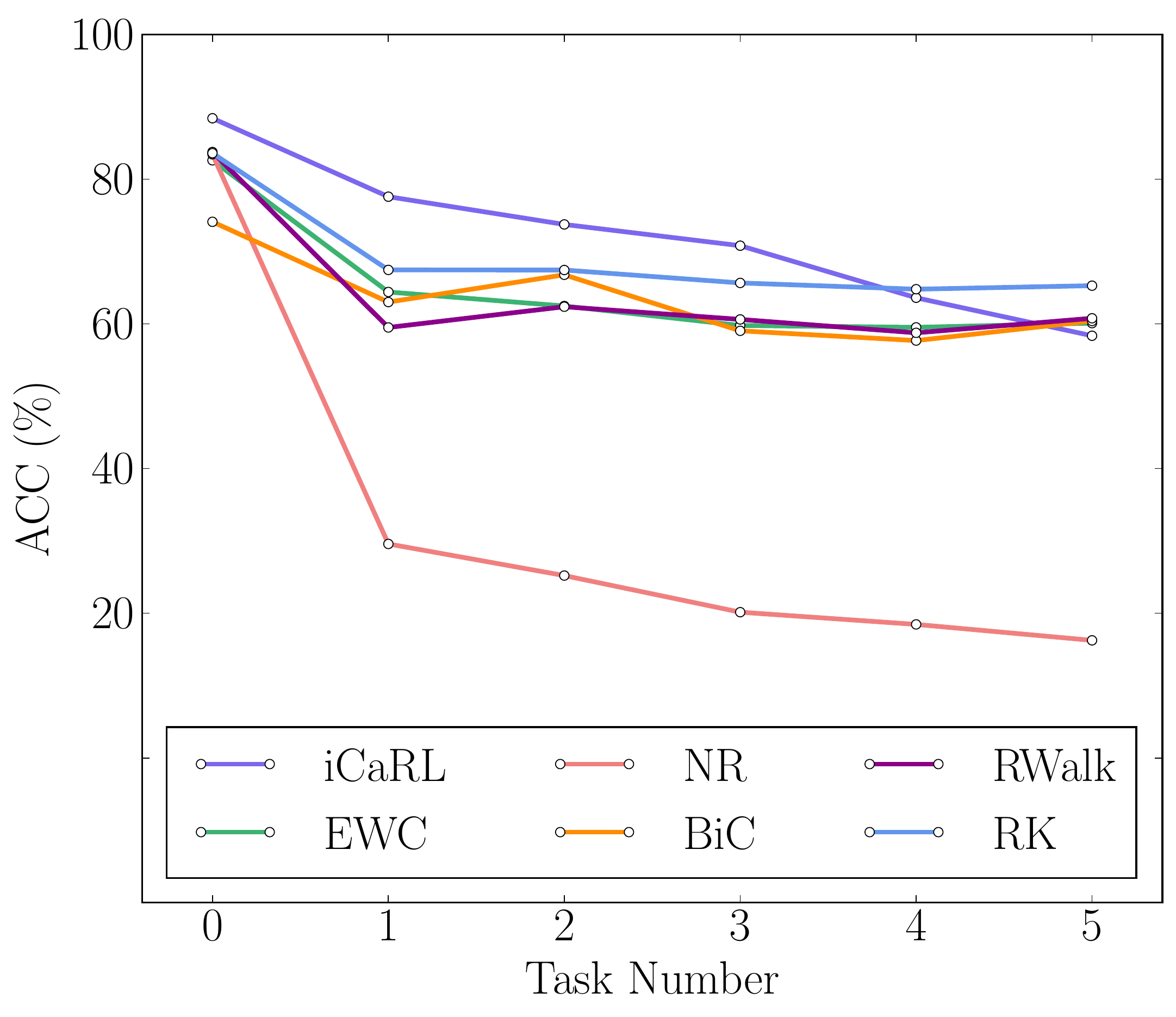}
  \centerline{(a) Memory size L = 300}
  \end{minipage}
\hfill
  \begin{minipage}[tbp]{0.49\linewidth}
  \centering
    \includegraphics[width=\linewidth]{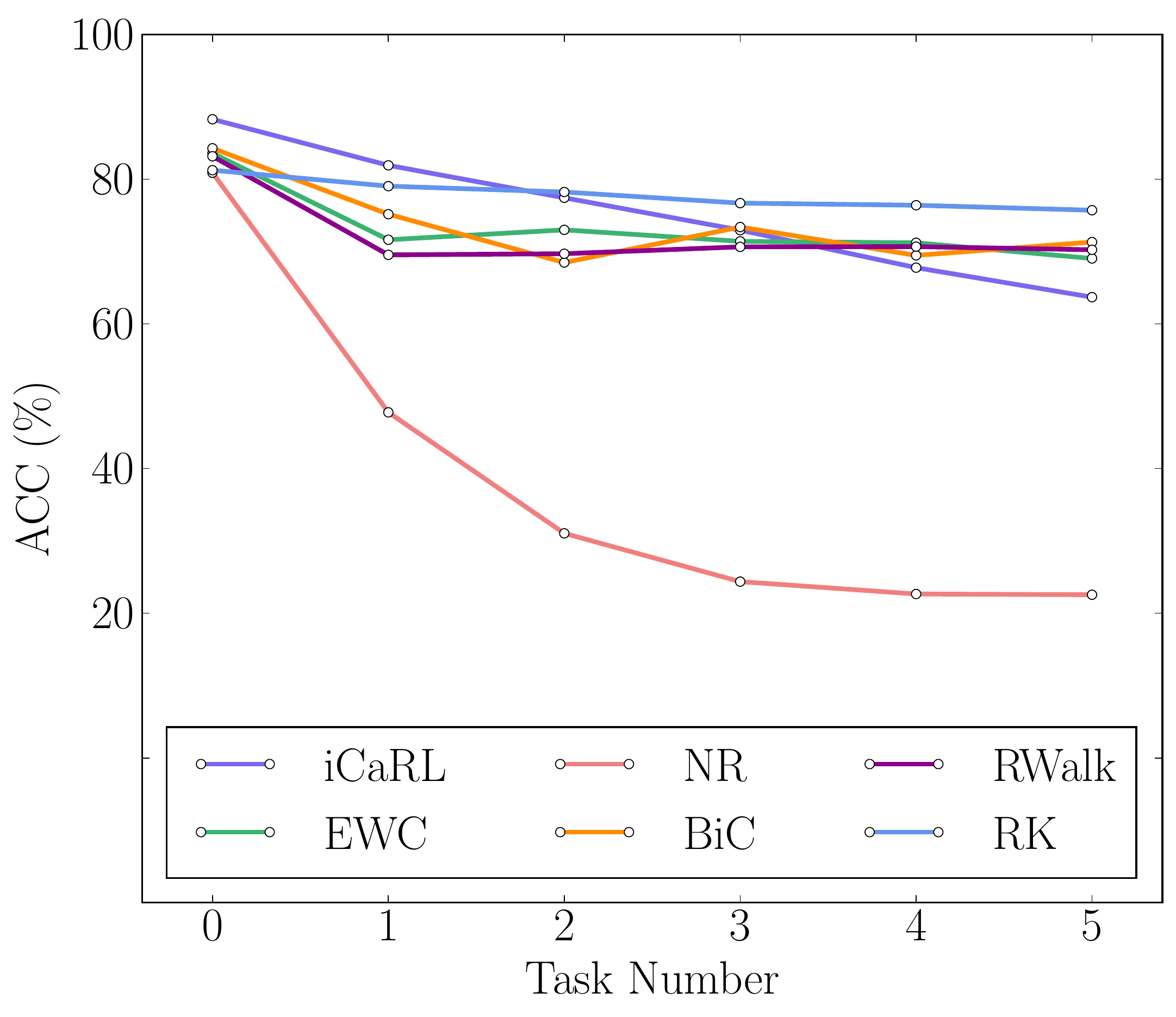}
  \centerline{(b) Memory size L = 500}    
  \end{minipage}
  \caption{The ACC (\%) in a comparative study of various memory size on the proposed RK approach and other competitive baselines.}
  \label{fig:size}
\end{figure}
\begin{table}[tb]
\centering
\caption{Average Accuracy (ACC) and Backward Transfer (BWT) in a comparative study of the proposed data augmentation. ``NoAugment" means no data augmentation is applied in the experiment. $L$ denotes the memory size for RK.}
\label{tab:da}
\resizebox{0.35\textwidth}{!}{%
\begin{tabular}{l|c|c}
\hline\hline
\textbf{Methods(L=500)} & \textbf{ACC($\uparrow$)} & \textbf{BWT($\uparrow$)} \\ \hline\hline
NoAugment      & 0.779          & -0.033         \\ \hline
SpecAugment \cite{park2019specaugment}    & 0.783          & \textbf{-0.006 }    \\ \hline

Mixup \cite{zhang2017mixup} & \textbf{0.828} & -0.036  \\ \hline\hline
\end{tabular}%
}
\end{table}
\subsubsection{Effect of various memory size on RK approach}
We then report the effect of various memory size (L=300 or 500) on RK approach, as shown in Figure \ref{fig:size}. We constrain all methods under the same memory size as that of the RK approach. We observe that even all approaches perform much better with increasing memory size (i.e., more available training data), the proposed RK approach still outperforms other methods. Furthermore, even with limited memory size (L=300), the proposed RK approach can obtain over 65.0\% ACC performances, much better than other methods.

\blfootnote{This research is supported by the National Research Foundation, Singapore under its AI Singapore Programme (AISG Award No: AISG-100E-2018-006).

The computational work for this article was partially performed on resources of the National Supercomputing Centre, Singapore (https://www.nscc.sg).}

\subsubsection{Effect of the data augmentation}
We also summarize the performances of the effects with two data augmentation methods on the proposed RK approach, as shown in Table \ref{tab:da}. We observe that all data augmentation methods improve the performances in terms of ACC. The best performances are achieved by the 'Mixup' data augmentation. We adopt the 'Mixup' data augmentation hereafter.

\begin{table}[tbp]
\centering
\caption{Accuracy and efficiency metrics in a comparative study of recent state-of-the-art training strategies. We adopt the TC-ResNet-8 model as testbed for all training strategies for fair comparison. Memory size $L$ is set to {[}500, 1500, 3000{]} in following experiments for RK.}
\label{tab:fn}
\resizebox{0.45\textwidth}{!}{%
\begin{tabular}{c|c|c|c|c}
\hline\hline
\textbf{Methods} & \textbf{ACC($\uparrow$)}   & \textbf{BWT($\uparrow$)}    & \textbf{Parameters} & \textbf{Memory} \\ \hline\hline
Fine-tune        & 0.262          & -0.372          & 64.48K              & 162.4M               \\ \hline
EWC              & 0.835          & -0.064         & 129.96K             & 162.4M               \\ \hline
PCL-KWS          & 0.836          & -0.041          & 406.9K              & 162.4M  
           \\ \hline
NR               & 0.560          & -0.163          & 64.48K              & 178.6M               \\ \hline
iCaRL            & 0.846          & -0.057          & 75.29K              & 178.6M               \\ \hline
BiC              & 0.793          & -0.085          & 64.48K                  & 178.6M               \\ \hline
RWalk            & 0.871          & -0.045          & 129.96K              & 178.6M    
             \\ \hline\hline
RK-500           & 0.828          & -0.036          & 129.96K              & \textbf{16.2M}       \\ \hline
RK-1500          & 0.887          & -0.012          & 129.96K              & 48.6M                \\ \hline
RK-3000          & \textbf{0.913} & \textbf{-0.012} & 129.96K              & 97.2M                \\ \hline\hline
Joint            & 0.940          & -               & 64.48K              & 1624.4M              \\ \hline
\end{tabular}%
}
\end{table}

\subsubsection{Benchmark against other competitive methods}
Table \ref{tab:fn} summarizes the comparison between the proposed RK and other competitive methods in terms of Average Accuracy (ACC), Backward Transfer (BWT), Parameters and Memory. The task number is set to 5. We observe that the proposed RK achieves the best performance with memory size $L$ of 3000. Comparing with the best baselines: RWalk, the proposed RK-3000 achieves 4.2\% absolute improvements in terms of Average Accuracy with fewer memory, which is closer to the upper-bound performances. Furthermore, even with only 16.2M training data, our approach RK-500 has comparable performance to other baseline methods, which is effective on edge devices. 

\section{Conclusions}
In this paper, we propose a novel diversity-aware class incremental learning method named Rainbow Keywords (RK) approach to avoid catastrophic forgetting with less memory. Experimental results show that the proposed RK approach achieves 4.2\% absolute improvement in terms of average accuracy over the best baseline. Ablation study also indicates that the proposed data augmentation and knowledge distillation loss are quite effective on edge devices.

% \section{Acknowledgement}

% This research is supported by the National Research Foundation, Singapore under its AI Singapore Programme (AISG Award No: AISG-100E-2018-006). The computational work for this article was partially performed on resources of the National Supercomputing Centre, Singapore (https://www.nscc.sg).  Sincerely thank Huaizheng Zhang and Yizheng Huang for the discussion and mentoring.

\clearpage

\input{interspeech_v11.bbl}

\bibliographystyle{IEEEtran}

% \bibliography{interspeech_v11}

% \begin{thebibliography}{9}
% \bibitem[1]{Davis80-COP}
%   S.\ B.\ Davis and P.\ Mermelstein,
%   ``Comparison of parametric representation for monosyllabic word recognition in continuously spoken sentences,''
%   \textit{IEEE Transactions on Acoustics, Speech and Signal Processing}, vol.~28, no.~4, pp.~357--366, 1980.
% \bibitem[2]{Rabiner89-ATO}
%   L.\ R.\ Rabiner,
%   ``A tutorial on hidden Markov models and selected applications in speech recognition,''
%   \textit{Proceedings of the IEEE}, vol.~77, no.~2, pp.~257-286, 1989.
% \bibitem[3]{Hastie09-TEO}
%   T.\ Hastie, R.\ Tibshirani, and J.\ Friedman,
%   \textit{The Elements of Statistical Learning -- Data Mining, Inference, and Prediction}.
%   New York: Springer, 2009.
% \bibitem[4]{YourName17-XXX}
%   F.\ Lastname1, F.\ Lastname2, and F.\ Lastname3,
%   ``Title of your INTERSPEECH 2022 publication,''
%   in \textit{Interspeech 2022 -- 23\textsuperscript{rd} Annual Conference of the International Speech Communication Association, September 18-22, Incheon, Korea, Proceedings, Proceedings}, 2022, pp.~100--104.
% \end{thebibliography}

\end{document}